\documentclass[times,aps,pre,twocolumn,byrevtex,amsmath,amssymb,floatfix,superscriptaddress,showpacs]{revtex4}  

\usepackage{graphicx} 
\usepackage{dcolumn} 
\usepackage{bm} 
\usepackage{times}
\usepackage{colordvi}
\usepackage{amsmath}
\usepackage{enumerate}
\usepackage{gensymb}

\begin{document}

\newcommand{\Fbox}[1]{\mbox{$\bigcirc\!\!\!\!${\scriptsize #1}$\;$}}


\title{Self-Assembly of Magnetic Spheres in Strong Homogeneous Magnetic Field}

\author{Ren\'e Messina}
\email{rene.messina@univ-lorraine.fr}
\affiliation{
 Laboratoire de Chimie et Physique - Approche Multi-Echelle des Milieux Complexes (LCP - A2MC)
 Institut de Chimie, Physique et Mat\'eriaux (ICPM),
 Universit\'e de Lorraine,
 1 Boulevard Arago,
 57070 Metz,
 France}

\author{Igor Stankovi\'{c}}
\email{igor.stankovic@ipb.ac.rs}
\affiliation{Scientific Computing
Laboratory, Institute of Physics Belgrade, University of Belgrade,
Pregrevica 118, 11080 Belgrade, Serbia}

\begin{abstract}

The self-assembly in two dimensions of spherical magnets in strong
magnetic field is addressed theoretically. 
It is shown that the attraction and assembly of parallel magnetic chains 
is the result of a delicate interplay of dipole-dipole interactions and short ranged excluded volume correlations. 
Minimal energy structures are obtained by numerical optimization procedure as well as
analytical considerations. For a small number of constitutive
magnets $N_{\rm tot}\leq26$, a straight chain is found to be stable. 
In the regime of larger $N_{\rm tot}\geq27$, the magnets form \textit{two touching} 
chains with equally long tails at both ends. 
We succeed to identify the transition from \textit{two} to \textit{three} touching chains at $N_{\rm tot}=129$.

\end{abstract}

\maketitle

\section{Introduction}

There are several reasons for strong and growing interest in self-assembled
structures of dipolar particles (i.e., with electric or magnetic dipoles). 
On the technological side, such systems have enormous potential  applications.
For instance, manufacturing of novel optical and stimuli-responsive materials
is based on self assembly of magnetic particles 
\cite{Zeng_Nature_2002,Wang_MaterialsToday_2013}.
On the other hand, dipolar particles and the the resulting phases 
can be well tuned by imposing an external field \cite{Blaaderen_Nature_2003,Malik_Langmuir_2012}. 
Assemblies of magnetic particles are known to produce a plethora
of one-, two-, and three dimensional objects (e.g., chains, rings,
and even tubes) 
\cite{Pal_PRE_2011,Messina_PhysRevE_2014,Messina_EPL_2015}.
From the perspective of biophysics, magnetic particles can be regarded as a model system
for probing the polar organization of microtubules 
\cite{Spoerke_Langmour_2008} or generating spontaneous helical superstructures 
\cite{Singh_Science_2014} reminiscent of DNA molecules. 

By essence, the dipole-dipole driving force for self-assembly is long range and
highly anisotropic (i.e., non-central pair potential) \cite{Kocbach_PhysicsEducation_2010,Vandewalle_NJP_2014} 
and therefore represents a formidable theoretical challenge. 
In this spirit, the pioneering theoretical
work of Jacobs and Beans~\cite{Jacobs_PhysRev_1955} and later that
of de Gennes and Pincus~\cite{deGennes_PhysKondMat_1970} 
about the microstructure of self-assembled (spherical)
magnets shed some light on the ordering mechanisms. 
More recently, microstructures of dipolar fluids have been thoroughly studied by computer
simulations~\cite{Prokopieva_PhysRevE_2009,Kantorovich_PRE_2012,Kaiser_PhysRevE_2015}
and in experiments~\cite{Heinrich_PRL_2011}. There, an important common
feature is the formation of chains  
\cite{Prokopieva_PhysRevE_2009,Kaiser_PhysRevE_2015}
and possibly in presence of an external
magnetic field \cite{Heinrich_PRL_2011,Kantorovich_PRE_2012}. 
With all that being said, only recently, the
ground state structures of magnetic spheres without external magnetic field have been
properly addressed in three dimensions 
\cite{Messina_PhysRevE_2014,Messina_PREReply_2015,Friedrich_PREComment_2015} 
as well as in two dimensions \cite{Messina_EPL_2015}. 

The goal of the present contribution is to tackle the fascinating problem
of self-assembly of magnets under strong magnetic field 
in a physically simple and transparent framework.
We explore (effective) interactions and assemblies of magnetic beads with \textit{parallel dipoles} 
(e.g., as obtained by a strong external magnetic field) and confined in two dimensions (e.g., by gravity).
We utilize two fully different routes to calculate the energy
minimum of the system: (i) genetic algorithm and (ii) direct
calculation and comparison of the energy of different configurations. 
The paper is organized as follows: 
In Sec. II we expose the magnetic chains Hamiltonian. 
Sec. III is devoted to to the analytical results dealing with the two-chain state. 
Phase diagram of self-assembled magnets obtained by numerical genetic algorithm is
discussed in sec. IV.   
Concluding remarks are provided in Sec. V. 

\section{Model}

\subsection{Pair interaction potential}

We begin by considering two magnetic hard spheres of diameter $d$
separated by a distance $r_{12} = |\vec r_2 - \vec r_1|>d$,
where $\vec r_1$ and $\vec r_2$ represent the position vectors of the centers
of particle 1 and particle 2, respectively.
These spherical magnets being also characterized by a magnetic moment ($\vec m_1, \vec m_2$),
the pair potential energy is dictated by:
\begin{equation}
\label{eq.U_pair} U(\vec r_{12}) = C \displaystyle \frac{1}{r_{12}^3}
\left[ \vec m_1 \cdot \vec m_2 - 3 \frac{(\vec m_1 \cdot \vec
r_{12}) (\vec m_2 \cdot \vec r_{12})  }{r_{12}^2} \right].
\end{equation}
%
where $C$ is  a constant that depends on the intervening medium
(eg, for vacuum $C=\frac{\mu_0}{4\pi}$ with $\mu_0$ being the vacuum permeability).

The approach we adopt in this work is based on the calculation of the magnetic energy of
various configurations of spheres in externally imposed magnetic
field $\vec B=B\vec e_x$ aligned with $x$-axis. We assume that the external
magnetic field is strong, i.e, $B\gg mC\frac{1}{d^3}$
(with $m := |\vec m_1| = |\vec m_2|$), so that all
dipole moments in the system are aligned with $\vec B$ and hence parallel to the $x$-axis,
i.e. $\vec m = m \vec e_x$.
In this limit of strong field, the pair potential \eqref{eq.U_pair} becomes
\begin{equation}
\label{eq.U_pair_B}
U(\vec r_{12}) = C \displaystyle \frac{m^2}{r_{12}^3}
\left[1  - 3 \frac{(x_2-x_1)^2}{r_{12}^2} \right].
\end{equation}
%

\subsection{Chains Hamiltonian}

It is convenient to introduce the energy scale defined
by $U_{\uparrow\uparrow} \equiv \frac{Cm^2}{d^3}$ that physically
represents the repulsive potential value for two parallel dipoles
at contact standing side by side as clearly suggested by the notation.
The dipoles attract if placed with head-to-tail (i.e., $\rightarrow \rightarrow $).
The latter promotes formation of one
dimensional chains consisting of many dipoles (i.e., $\rightarrow
\rightarrow \dots \rightarrow$).
The reduced total potential energy of interaction of a system consisting of two chains made up of $N_1$ and $N_2$
particles, $U^{tot}_{N_1N_2}$, can be written as
\begin{equation}
\label{eq.U_rescaled}
U^{\rm tot}_{N_1N_2} = \displaystyle
\frac{1}{2}
\mathop{\sum_{i,j=1}^{N_1+N_2}}_{i\neq j}
\frac{U(\vec r_{ij})}{U_{\uparrow\uparrow}} \quad (r_{ij} \geq d).
\end{equation}
The sum in Eq. \eqref{eq.U_rescaled} can be separated into three terms,
\begin{equation}
\label{eq.U_decompos}
U^{tot}_{N_1N_2}=U^{\rm 1c}_{N_{\rm 1}}+U^{\rm 1c}_{N_{\rm
2}}+U^{cc}_{N_{\rm 1}N_{\rm 2}},
\end{equation}
where $U^{\rm 1c}_{N_{\rm 1,2}}$ are the reduced \textit{intra-chain}
contributions to the total energy, whereas $U^{\rm cc}_{N_{\rm 1}N_{\rm
2}}$ is the \textit{inter-chain} (or cross chain) contribution to the total energy.

More explicitly, the reduced intra-chain energy $U^{1c}_{N}$ is given by
\begin{equation}
\label{eq.u1c} U^{\rm 1c}_{N} = \displaystyle
\sum^{N-1}_{i=1}\sum^{N}_{j=i+1}\frac{U(\vec
r_{ij})}{U_{\uparrow\uparrow}} \quad (r_{ij} \geq d)
\end{equation}
respecting the non-overlapping conditions.
Thereby, the expression for $U^{\rm 1c}_{N}$ in reduced units is merely given by
%
\begin{equation}
\label{eq.energy_intrachain}
U^{\rm 1c}_{N} = -\sum_{i=1}^{N-1}\sum_{j=i+1}^{N}\frac{2}{(j-i)^3}
= -2\sum_{i=1}^{N-1}\frac{N-i}{i^3}.
\end{equation}
%
This always negative energy in Eq. \eqref{eq.energy_intrachain}
can be seen as the \textit{cohesion} energy of a magnetic chain.
%
On the other hand, the inter-chain term  $U^{cc}_{N_{\rm 1}N_{\rm 2}}$
in Eq. \eqref{eq.U_decompos} can be either positive or negative depending
on the relative $x$-shift of the two chains.
Assume chain 1 has its first bead at $x_1^{(1)}$
so that its last one is at $x_1^{(N_1)}=x_1^{(1)}+N_1d$.
Given the symmetry of the system, chain 2 position is then fully specified
by the position of its first bead $x_2^{(1)}= x_1^{(1)} + \delta_xd$,
with $\delta_x$ representing the relative $x$-shift of the two chains.
The relative position of two beads $i$ and $j$ belonging to chain 1 and chain 2, respectively,
can be written as $\vec r_{ij}/d=(j-i+\delta_x,\delta_y)$, where $\delta_y$ is the relative (reduced)
$y$-position of the two chains.
We then arrive at the simple expression for cross chain interaction energy

\begin{eqnarray}
\nonumber
U^{\rm cc}_{N_{\rm 1}N_{\rm 2}}(\delta_x,\delta_y) & = &
\sum^{N_{\rm 2}}_{i=1}\sum^{N_{\rm
1}}_{j=1}\left\{\frac{1}{[(j - i + \delta_x)^2 +
\delta_y^2]^\frac{3}{2}} \right. \\
&&
- \left. \frac{3(j - i + \delta_x)^2}{[(j - i + \delta_x)^2 +
\delta_y^2]^\frac{5}{2}}\right\}.
\label{eq.Ucross_N1N2}
\end{eqnarray}
%
The effective force between these two chains in the direction perpendicular to the
external magnetic field is then given by
$F_y^{cc} = -\nabla_{\delta_y} U^{cc}_{N_{\rm 1}N_{\rm 2}}(\delta_x,\delta_y)$.
Hence, when talking about attraction vs repulsion, it is the sign of
that effective force $F_y^{cc}$
(or $F_x^{cc}= -\nabla_{\delta_x} U^{cc}_{N_{\rm 1}N_{\rm 2}}$
for the component in the field direction) that will matter.

\begin{figure}
\includegraphics[width = 8.5 cm]{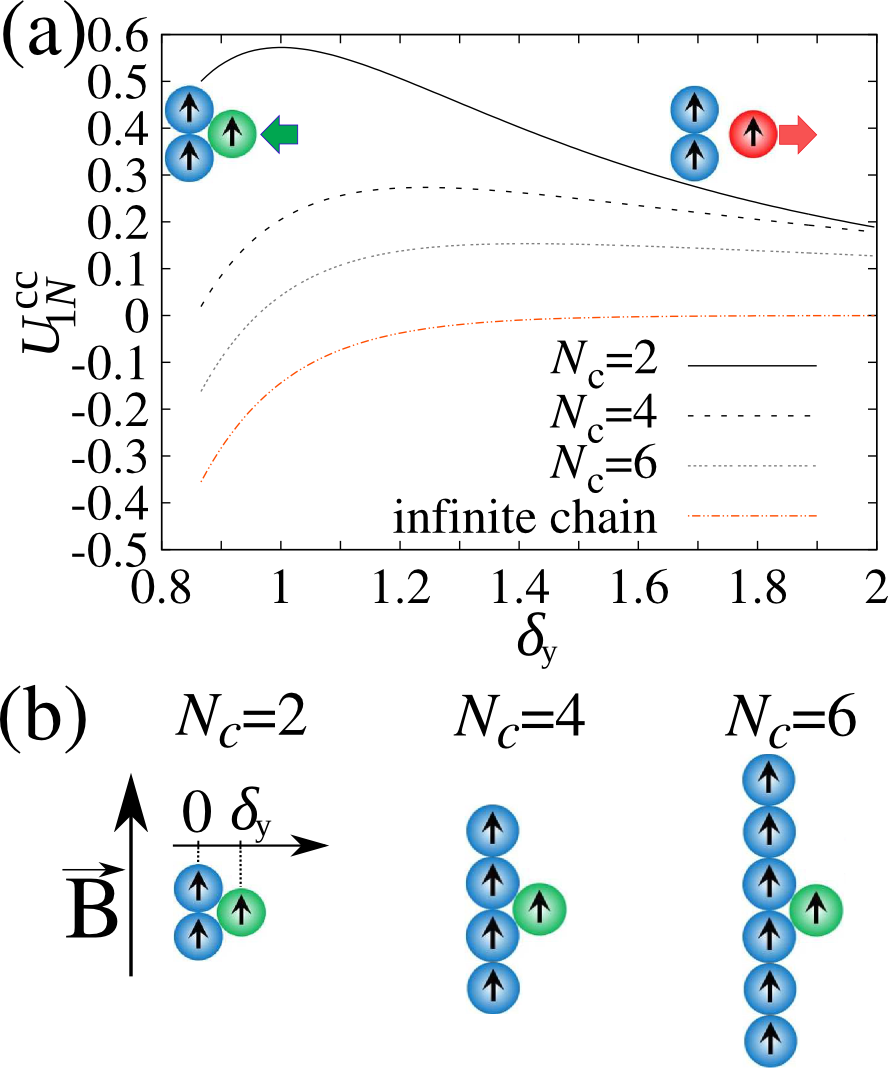}
\caption{
(a) Reduced cross energy profile of a a single magnet interacting with a chain made up of $N_c$ magnetic beads.
The center of the single magnet is located at a $\delta_y$ distance from the chain and always at the
$x$-mid-height of the latter.
(b) Sketch at contact for $N_c=2,4,6$ where $\delta_y=\sqrt{3}/2$.}
\label{fig:nipple}
\end{figure}

\section{Interaction of two chains}

\subsection{Chain-dipole interaction}
\label{sec:nipple}

It is instructive to first consider the interaction between a
chain and a single magnet. Thereby, the most elementary situation
consists of a single magnet (i.e., $N_1=1$) interacting with a
dimer (i.e., $N_2=2$).
An important configuration is that corresponding to a triangle,
see inset in Fig. \ref{fig:nipple} for $N=2$, since this is also
the local configuration of an infinite triangular lattice.
Thereby, the (equilateral) triangle configuration energy is merely
given by $U^{cc}_{12}(\delta_x=1/2,\delta_y=\sqrt3/2)=1/2$, see
also Fig. \ref{fig:nipple}. The positive value indicates that
there is a strong energy penalty upon assembling a magnetic dimer
and a single magnet together into a triangle from infinite
relative separation. 
This feature is fully consistent with the geometrical idea that 
$60^{\circ}$ is larger than the magic angle ($54.7^{\circ}$) 
\cite{Kocbach_PhysicsEducation_2010}. 
Interestingly, upon slightly separating the
two objects (keeping $\delta_x=1/2$), there is a slight increase
in energy proving an effective attraction, see Fig.
\ref{fig:nipple}. At large separation, one recovers the behavior
of two  point-like dipoles leading to a typical repulsion scaling
as $1/\delta_y^3$.

When increasing the chain size by the same amount on both ends,
the energy at contact for a flat T-shaped configuration is
significantly decreased, see Fig. \ref{fig:nipple}. This is merely
due to the attractive terms stemming from the interaction between
the further dipoles of the chain with the single aside magnet.
This energy at contact becomes asymptotically
$U^{cc}_{1\infty}(\delta_y=\sqrt3/2) \simeq -0.356$ for an
infinite chain. Concomitantly, the energy barrier upon approaching
a magnet from infinity vanishes when the chain gets very large,
see Fig. \ref{fig:nipple}. Thus, chain size qualitatively matters,
and peripheral dipoles along the chain screen the repulsion
between the aside magnet and its first neighbours. At small
$\delta_y$-separation, deep minima are attained (see Fig.
\ref{fig:nipple}) showing that the interaction strength with
satelite dipoles overcompensates that with first neighbours.


\subsection{Two-chain state}

\subsubsection{Like sized chains}

We now would like  to provide a more quantitative analysis of
interaction of two equally long chains. Taking advantage of the
symmetry  the cross energy expression with respect to indices when
$N_1=N_2=N$ in Eq. \eqref{eq.Ucross_N1N2}, we find:
\begin{eqnarray}
\label{eq.Ucross_NN}
\nonumber
U^{\rm CC}_{NN}(\delta_x,\delta_y) & = & \sum^{N-1}_{i=-N+1}(N-|i|)
\left\{\frac{1}{[(i + \delta_x)^2 + \delta_y^2]^\frac{3}{2}} \right. \\
&&
   \left. - \frac{3(i + \delta_x)^2}{[(i + \delta_x)^2 +
   \delta_y^2]^\frac{5}{2}}\right\}.
\end{eqnarray}

\begin{figure}
\includegraphics[width = 8.5 cm]{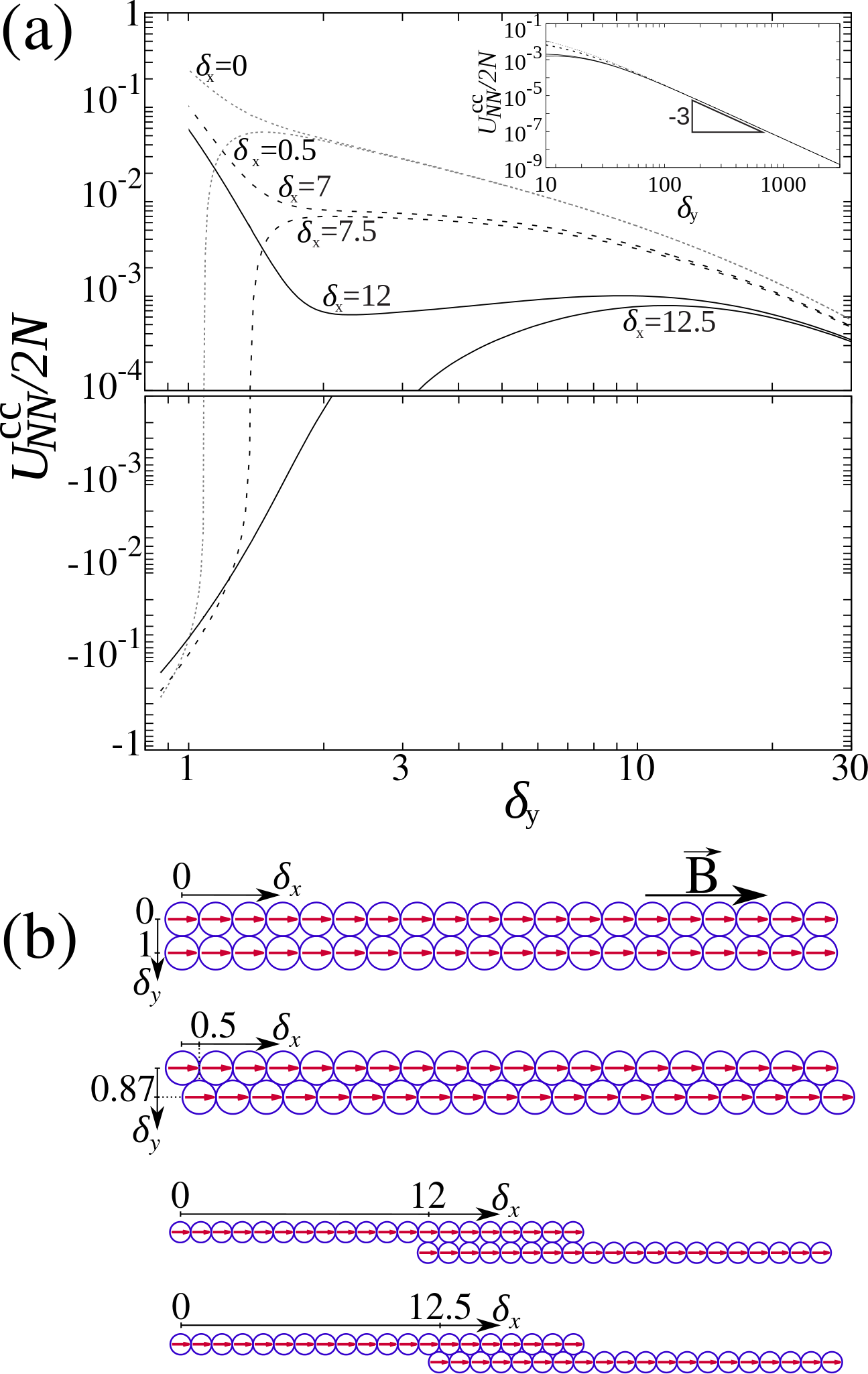}
\caption{(a) Interchain interaction energy (per particle) profile as a function
of the reduced transverse chain-chain $\delta_y$-separation for different values
of the lateral chain-chain reduced shift $\delta_x$. Notice the breaking in the energy-axis
allowing logarithmic scale for positive and negative values.
(b) Microstructures of chains at contact for different values of $\delta_x$ oriented in the
magnetic field $\vec B$ direction.
}
\label{fig:Ucross_N20}
\end{figure}

The profiles of the interaction potential  of two chains
with $N=20$ particles can be found in Fig.~\ref{fig:Ucross_N20}.
In this scenario, we want to illustrate the crucial effect of
\textit{positional lateral correlations}.
As expected, the strongest repulsion occurs for chains standing
exactly face to face with zero lateral shift (i.e., $\delta_x=0$),
see Fig.~\ref{fig:Ucross_N20}. A small shift amount, here
$\delta_x=1/2$ (see Fig.~\ref{fig:Ucross_N20}), drastically
changes the situation where now effective attraction takes place
at short transverse $\delta_y$-separation. Differently said,
self-assembly with $\delta_x=1/2$ is now favored where the lowest
energy is obtained at contact with local high packing,
\footnote{We have carefully checked that the deepest minimum for $0 < \delta_x <1$ occurs at contact and
for $\delta_x=0.5$ as intuitively expected.}
see Fig.~\ref{fig:Ucross_N20}.
Upon increasing the relative lateral shift $\delta_x$ between the
chains, a similar behavior is observed. More specifically, shifted
chains with beads standing exactly face to face (i.e., for
$\delta_x$ assuming integer values, say $n$) always lead to higher
energy $\delta_y$-profiles than with $\delta_x=n+1/2$, especially
near contact, see Fig.~\ref{fig:Ucross_N20}(b). Besides, the near
field repulsion observed for integer values of $\delta_x$ switches
to attraction if particles of one chain are in the
$x$-middle-point positions of the other chain ones (i.e., for
$\delta_x=0.5,5.5,$ and $12.5$), see Fig.~\ref{fig:Ucross_N20}.

In the far field limit (i.e., $\delta_y/N \gg 1$), magnetic chains
should behave as super dipoles, i.e., point-like dipoles
with strength $Nm$. Indeed, at sufficiently large chain-chain
distance repulsion sets in with a $~1/\delta_y^3$ power law
dependence on distance, see inset in Fig.~\ref{fig:Ucross_N20}(a).
The point where near field attraction switches into repulsion can
be understood as crossover between far and near field behavior.
As a matter of fact, lateral correlations between discrete finite magnets
are essential to promote effective attraction between facing chains within distances
that are of the order of the chain itself (i.e., $\delta_y/N \lesssim 1$).

%

\begin{figure}
\includegraphics[width = 8.5 cm]{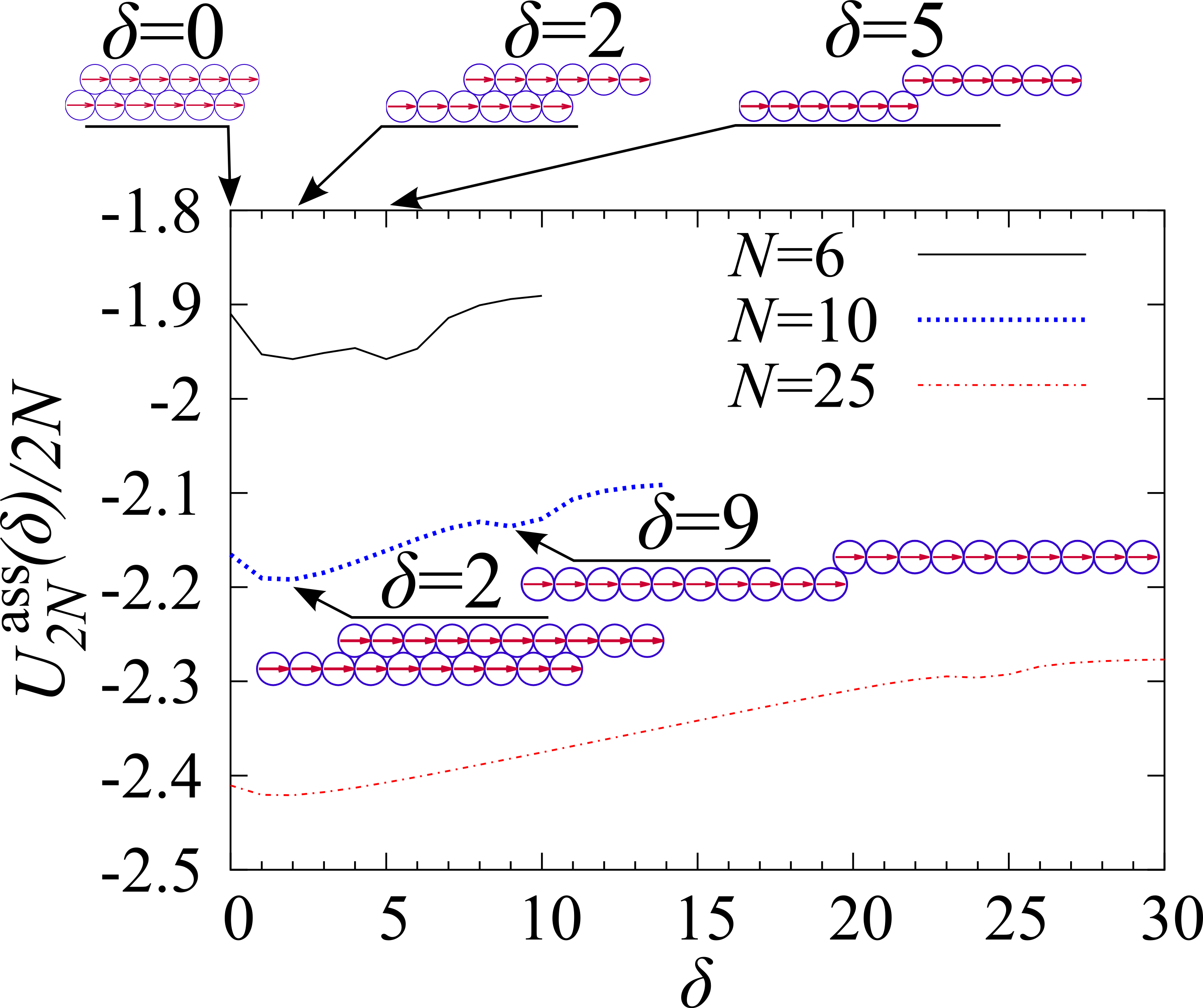}
\caption{
Total reduced energy per particle, $U_{NN}^{\rm ass} (\delta)/2N$,
as a function of the relative lateral displacement $\delta$.
Microstructures correponding to minima are shown for $N=6,10$.
Note that only the range $\delta \leq N-1$ corresponds to touching chains
as clearly illustrated by the depicted microstructures.
}
\label{fig:Utot_tails_equal_chain_size}
\end{figure}

To isolate the effects of lateral displacement along the chain axis (or equivalently the external magnetic field),
we consider two touching  equally long chains exhibiting locally a densely packed triangular structure,
see microstructures in Fig. \ref{fig:Utot_tails_equal_chain_size}.
It is useful in this context to introduce integer values of displaced beads via the relation
$\delta_x=\delta+1/2$.
Here the value of $\delta$ tells about the tail length, i. e.,  the number of beads of one chain end that
are not in contact with the other chain, see Fig. \ref{fig:Utot_tails_equal_chain_size}
for illustrative configurations with
$\delta=0,2,5,9$.
The total energy of our two assembled chains,
$U_{2N}^{\rm ass} \equiv U^{\rm tot}_{NN}(\delta+1/2,\sqrt{3}/2)$,
can be written according to Eq. \eqref{eq.U_decompos}
as a sum of (i) twice the single chain cohesion energy $2U^{\rm 1c}$  and (ii)
the cross chain energy $U^{\rm cc}$:
\begin{equation}
\label{eq.energy_tot_2N}
U_{2N}^{\rm ass} (\delta) = 2U^{\rm 1c}_{N}+U^{\rm cc}_{NN}(\delta+1/2,\sqrt{3}/2).
\end{equation}
In a similar way that Eq. \eqref{eq.energy_intrachain} has been derived,
regrouping of repeating terms in the double sum [see Eq. \eqref{eq.Ucross_N1N2}] involved in
$U^{\rm cc}_{NN}(\delta+1/2,\sqrt{3}/2)$ entering Eq.\eqref{eq.energy_tot_2N}
leads to a simple single sum expression given by
\begin{eqnarray}
\label{energy}
\nonumber
&& U^{\rm cc}_{NN}(\delta + 1/2,\sqrt{3}/2)  =
\sum_{i=-N+1}^{N-1}(N-|i|) \\
&& \left\{\frac{1}{[(i+\delta+\frac{1}{2})^2+\frac{3}{4}]^\frac{3}{2}}
-\frac{3(i+\delta+\frac{1}{2})^2}{[(i+\delta+\frac{1}{2})^2+\frac{3}{4}]^\frac{5}{2}}
\right\}.
\end{eqnarray}
Profiles of the total energy per particle, $U_{NN}^{\rm ass} (\delta)/2N$, as a function
of the relative lateral displacement $\delta$
are displayed in Fig. \ref{fig:Utot_tails_equal_chain_size} for three chain lengths ($N=6,10,26$).
For short chains (here $N=6$) two minima are setting in, see Fig. \ref{fig:Utot_tails_equal_chain_size}.
The first and \textit{lowest} minimum occurs at $\delta=2$.
Its origin is a subtle balance between (i) a positive contribution
due to first neighbor interactions and (ii) a negative contribution
due to distant neighbor interactions.
These two mechanisms were clearly identified in Sec. \ref{sec:nipple}, see also Fig. \ref{fig:nipple}.
Since at $\delta=5$ the chains do not overlap (only touching ends),
the second observed minimum is purely a result of distant neighbour interactions,
see \ref{fig:Utot_tails_equal_chain_size}.
As the chain length is increased ($N=10,26$) these features pesrsit:
(i) always a lowest minima at $\delta=2$ and (ii)
a second minimum at chain-chain separation (i.e., $\delta=N-1$),
see Fig. \ref{fig:Utot_tails_equal_chain_size}.

\begin{figure}
\includegraphics[width = 8.5 cm]{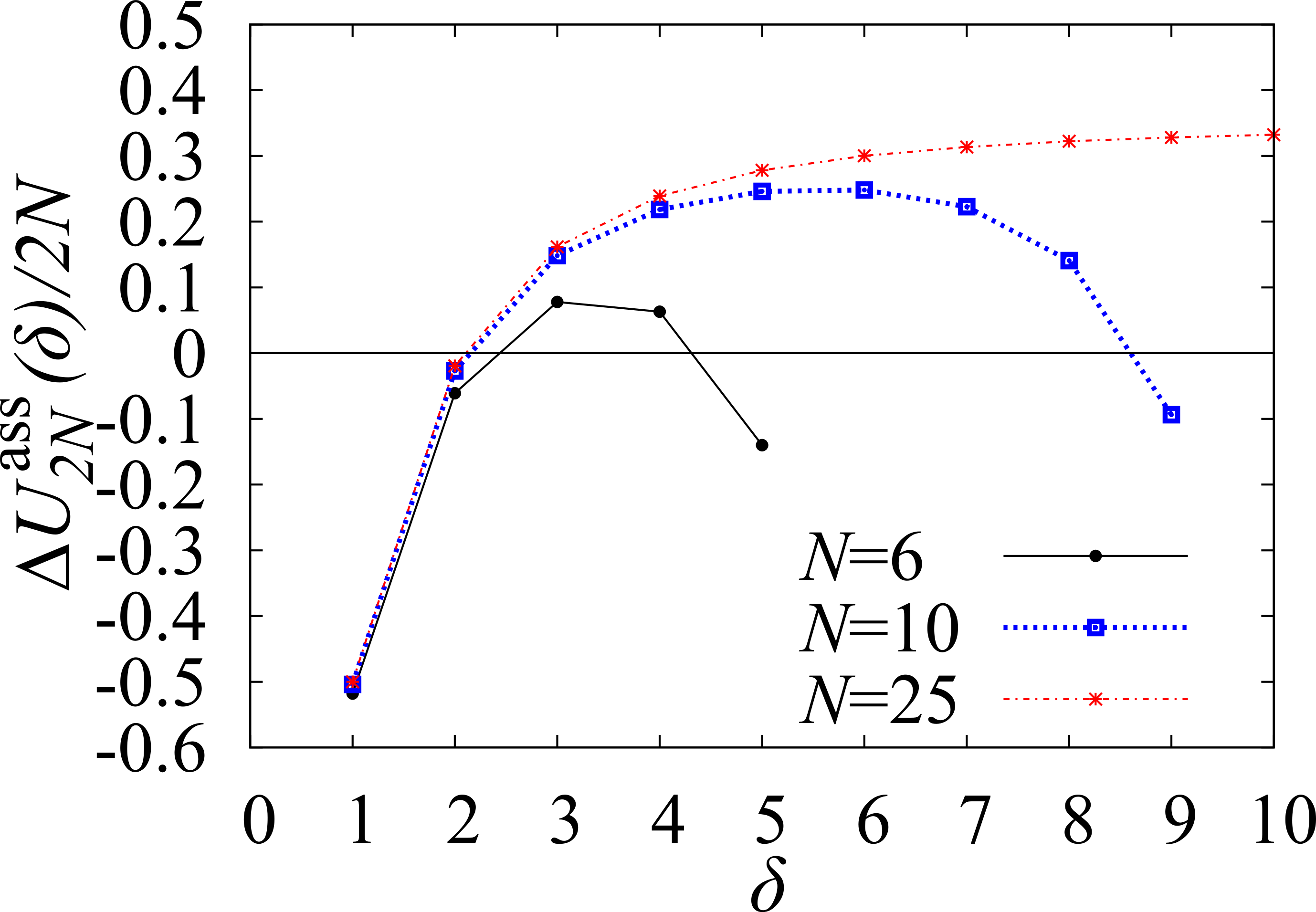}
\caption{
Profiles of the discrete energy variation $\Delta U_{NN}^{\rm ass} (\delta)$ as a function
of the relative lateral displacement $\delta$ for finite different values of $N$.
Only the range $\delta \leq N-1$ corresponding to interchain touching beads is shown.
}
\label{fig:Delta_Utot_tails_equal_chain_size}
\end{figure}

At this point, we would like to answer the following intriguing question:
Does the particular $\delta=2$ shift value  always correspond to the ground state for two like sized chains?
In other words, when two identical magnetic chains self-assemble in their ground state,
do they always exhibit three-bead long tails at both ends?
To rationalize this striking finding, we consider the following observable
\begin{equation}
\label{eq.def_energy change_NN}
\Delta U_{2N}^{\rm ass} (\delta) =
U_{2N}^{\rm ass} (\delta) -
U_{2N}^{\rm ass} (\delta-1)
\end{equation}
which is merely the energy difference of two states  at
$\delta$ and $\delta-1$, respectively.
\footnote{$\Delta U_{NN}^{\rm ass} (\delta)$ can also be seen as the backward discrete
derivative of $U_{NN}^{\rm ass}$ with respect to $\delta$.}
Since the cohesion energy is unchanged at prescribed chain size $N$,
the discrete energy variation $\Delta U_{2N}^{\rm ass} (\delta)$
given by Eq. \eqref{eq.def_energy change_NN} reads
\begin{eqnarray}
\label{eq.energy change_NN}
\nonumber
&&\Delta  U_{2N}^{\rm ass}  (\delta)  = \\
\nonumber
&&
\sum_{i=N-\delta}^{N+\delta+1}
\left\{\frac{1}{[(i - \frac{1}{2})^2 + \frac{3}{4}]^\frac{3}{2}}
      -\frac{3 (i- \frac{1}{2})^2}{[(i - \frac{1}{2})^2 + \frac{3}{4}]^\frac{5}{2}}
      \right\}  \\
&& - \sum_{i=-\delta}^{\delta+1}\left\{\frac{1}{[(i - \frac{1}{2})^2 +
\frac{3}{4}]^\frac{3}{2}}-\frac{3 (i- \frac{1}{2})^2}{[(i -
\frac{1}{2})^2 + \frac{3}{4}]^\frac{5}{2}}\right\}.
\end{eqnarray}
Profiles of  $\Delta U_{2N}^{\rm ass} (\delta)$ are shown in
Fig.~\ref{fig:Delta_Utot_tails_equal_chain_size} for finite $N$.
It clearly illustrates the behavior of  $\Delta U_{2N}^{\rm ass} (\delta)$
with respect to $\delta$ and $N$.
More specifically,  for overlapping chains (i.e., $\delta \leq N-1$),
the energy variation $\Delta U_{2N}^{\rm ass} (\delta)$ increases
at given $\delta$ with growing chain size $N$, see Fig.~\ref{fig:Delta_Utot_tails_equal_chain_size}.
Moreover, $\Delta U_{2N}^{\rm ass} (\delta)$ assume positive values for $2 < \delta < N-1$.
For large $N$, the second term in Eq. \eqref{eq.energy change_NN} converges to zero, so that the
behavior of $\Delta U_{2N}^{\rm ass} (\delta)$ is solely dictated by the first term.
A straightforward summation leads to
$\lim_{N \to \infty} \Delta U_{2N}^{\rm ass} (\delta=2) \simeq -0.019<0$
and $\lim_{N \to \infty} \Delta U_{2N}^{\rm ass} (\delta=3) \simeq +0.162<0$.
This demonstrates that $\delta=2$ is indeed the ground state of two assembled like sized chains.


\subsubsection{Unlike sized chains}

\begin{figure}
\includegraphics[width = 8.5 cm]{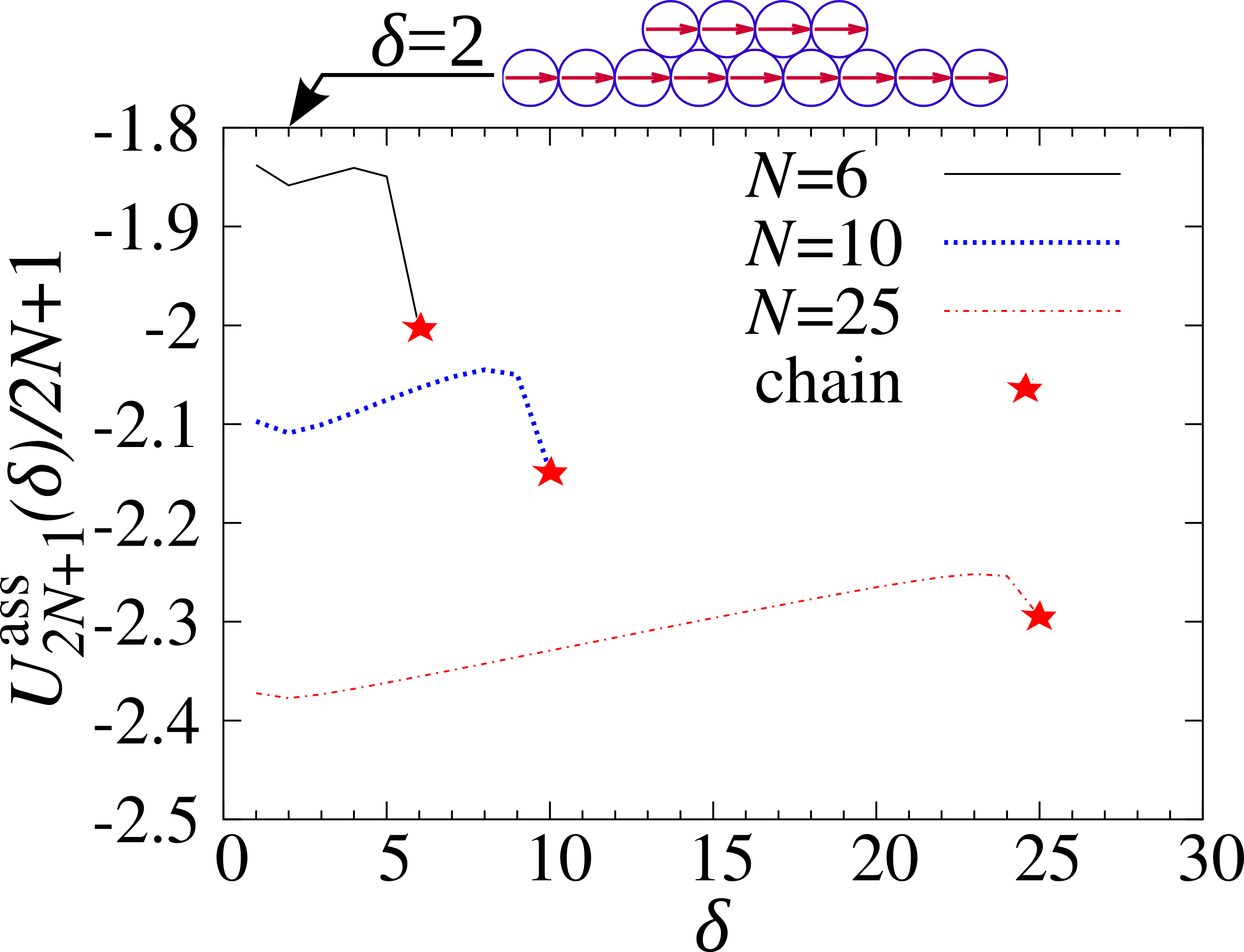}
\caption{
Total reduced energy per particle, $U_{2N+1}^{\rm ass} (\delta)/(2N+1)$,
as a function of the relative lateral displacement $\delta$.
The microstructure corresponding to the ground state (in the imposed two-chain regime)
with $N=6$ for $\delta=2$ is shown at the top.
The special case of $\delta=N$ corresponding to a \textit{single chain} made up of $2N+1$
beads is denoted by a star.
}
\label{fig:energy_ass_2N+1}
\end{figure}

In case of uneven total number of particles ($2N+1$), we pay attention
to the important situation of two symmetrically placed chains with $N+\delta+1$ and
$N-\delta$ beads, see Fig. \ref{fig:energy_ass_2N+1} for an illustration with $\delta = 2$.
\footnote{
Using a genetic algorithm in the next section it
will be confirmed that ground state
configurations (i.e., with minimal energy) belong indeed to this family of structures.
}
The energy of this system denoted by $U_{2N+1}^{\rm ass}(\delta)$ is given by
\begin{eqnarray}
\label{eq:U_2N+1^ass}
\nonumber
U_{2N+1}^{\rm ass}(\delta) & =  & U^{\rm 1c}_{N-\delta}+U^{\rm 1c}_{N+\delta+1}\\
&&
+U^{\rm cc}_{N-\delta,N+\delta+1}(\delta+1/2,\sqrt{3}/2),
\end{eqnarray}
%
where the cross chain term reads
\begin{eqnarray}
\label{eq:U_cross_2N+1}
\nonumber
&&U^{\rm cc}_{N-\delta,N+\delta+1}(\delta+1/2,\sqrt{3}/2) = \\ \nonumber
&& 2\sum_{i=0}^{N-1}{\rm min}\left( N-i,N - \delta \right) \\
&& \times \left\{\frac{1}{[(i+\frac{1}{2})^2+\frac{3}{4}]^\frac{3}{2}}
-\frac{3(i+\frac{1}{2})^2}{[(i+\frac{1}{2})^2+\frac{3}{4}]^\frac{5}{2}}
\right\},
\end{eqnarray}
and the intra-chain terms $U^{\rm 1c}_{N-\delta}$ and $U^{\rm 1c}_{N+\delta+1}$ are specified by Eq. \eqref{eq.energy_intrachain}.
Typical relevant profiles of  $\Delta U_{2N+1}^{\rm ass} (\delta)$ as given by Eq. \eqref{eq:U_2N+1^ass}
are shown in Fig.~\ref{fig:Delta_Utot_tails_equal_chain_size} for finite $N$.
Again, the value $\delta=2$ plays a special role in the energy minimum of assembled chains,
see Fig. \ref{fig:energy_ass_2N+1}.
It can be emphasized that,  in the regime of touching chains (i.e., when $\delta \leq N-1$),
a \textit{single} minimum appears and always at $\delta=2$!, see Fig. \ref{fig:energy_ass_2N+1}.
It is interesting to notice that for a given number of total
particles $2N+1$, the two-chain structure beats energetically the
single-chain one only when $N$ is large enough, see Fig.
\ref{fig:energy_ass_2N+1}. \footnote{ A similar conclusion could
be drawn for like sized chains, see Fig.
\ref{fig:Utot_tails_equal_chain_size}. }

\begin{figure}
\includegraphics[width = 8.5 cm]{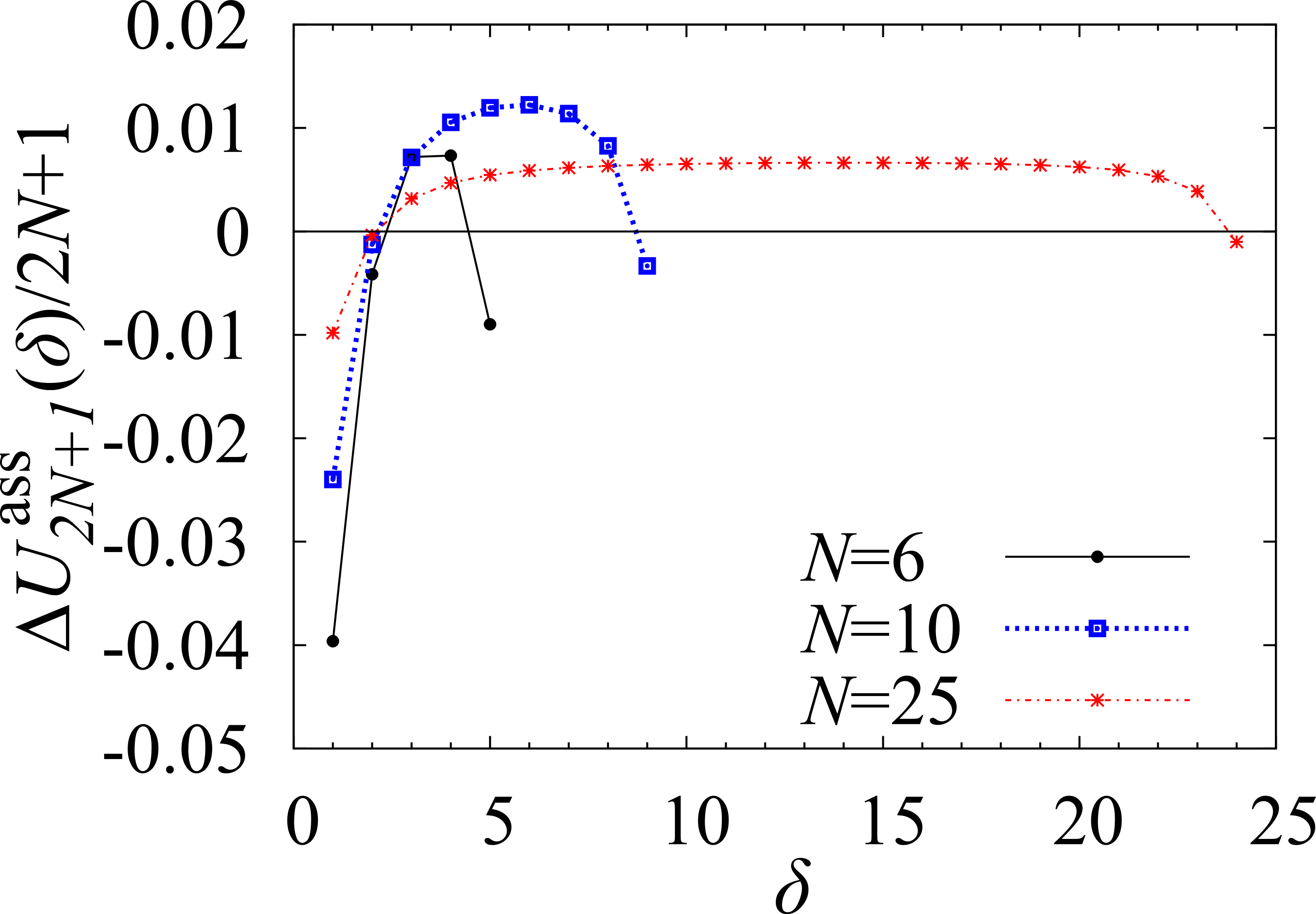}
\caption{
Profiles of the discrete energy variation $\Delta U_{2N+1}^{\rm ass} (\delta)$ as a function
of the relative lateral displacement $\delta$ for different values of $N$.
Only the range $\delta \leq N-1$ corresponding to two-chain states is shown, i.e.
the single chain configuration ($\delta = N$) is omitted.
}
\label{fig:dU_2N+1}
\end{figure}
In the same spirit of Eq. \eqref{eq.def_energy change_NN},  we define
the discrete energy change $\Delta U_{2N+1}^{\rm ass}(\delta)$ upon variation
of lateral displacement $\delta$ as
$\Delta U_{2N+1}^{\rm ass}(\delta) = U_{2N+1}^{\rm ass}(\delta) - U_{2N+1}^{\rm ass}(\delta-1)$.
It is a simple matter to show that this quantity reads
\begin{eqnarray}
\label{eq:dU_2N+1}
\nonumber
&& \Delta U_{2N+1}^{\rm ass}(\delta)  = -2 \sum_{i=N-\delta+1}^{N+\delta} \frac{1}{i^3}  \\
&& -2\sum_{i=0}^{\delta-1}
\left\{\frac{1}{[(i+\frac{1}{2})^2+\frac{3}{4}]^\frac{3}{2}}
-\frac{3(i+\frac{1}{2})^2}{[(i+\frac{1}{2})^2+\frac{3}{4}]^\frac{5}{2}} \right\},
\end{eqnarray}
where the first sum in Eq. \eqref{eq:dU_2N+1} stems from
intrachain interactions, and the second one from cross
chain interactions. The behavior of $\Delta U_{2N+1}^{\rm ass}(\delta)$ 
is sketched in Fig. \ref{fig:dU_2N+1}. We observe
that the energy difference changes sign from negative to
positive around $\delta=2$, which indicates a minimum in energy
there, see Fig. \ref{fig:dU_2N+1}. This minimum will also persist
at large $N$. Indeed, we learn from Eq. \eqref{eq:dU_2N+1} that
for large $N$ only the interchain correlations will
survive. In details, a straightforward summation leads to $\lim_{N \to \infty}
\Delta U_{2N+1}^{\rm ass} (\delta=2) \simeq -0.0188<0$ and
$\lim_{N \to \infty} \Delta U_{2N+1}^{\rm ass} (\delta=3) \simeq
0.1623>0$. Therefore, we can conclude that two symmetrically
placed touching chains of length $N+3$ and $N-2$ represent an
energy minimum for any $N$ larger than two.

\subsubsection{Summary}

Two chains self-assemble in a minimal energy configuration by
building short \textit{tails of two and a half beads} at both ends. This
striking and relevant feature is fully consistent with the picture
of a single magnet interacting with a chain, see Fig.
\ref{fig:nipple}. Thereby, it was indeed shown that the
interaction energy becomes \textit{negative} when the chain has at
least six beads ($N_c=6$ in Fig. \ref{fig:nipple} corresponding to $N_{tot}=2N+1=7$ with $N=3$), 
corresponding exactly to two tails of two and a half beads with respect to the touching
single magnet. In the next section, we will look for the overall
ground state at prescribed number of magnets $N_{tot}$, and
especially locate the regime of two-chain assemblies.

\begin{figure}
\includegraphics[width = 8.5 cm]{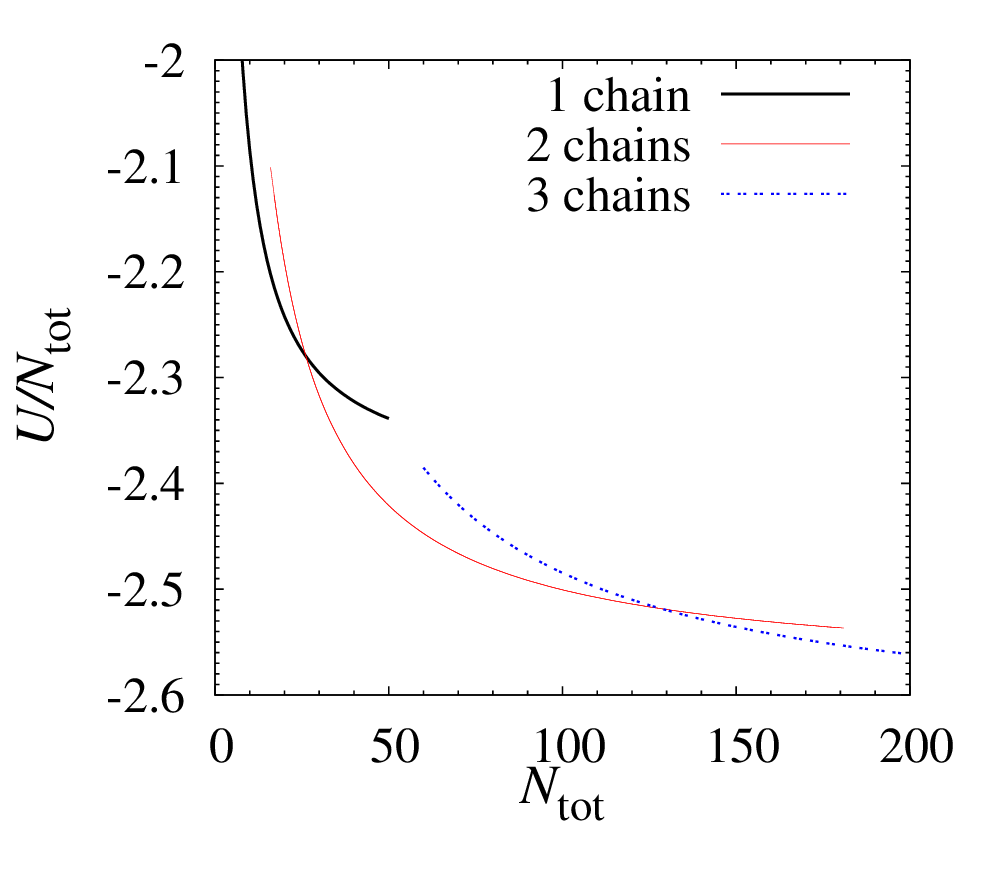}
\caption{Reduced energy per particle profiles as a function of the
total number of particles $N_{\rm tot}$. Ground states are as
follows: one chain ($2 \leq N_{\rm tot} \leq 26$), two chains ($27
\leq N_{\rm tot} \leq 128$), and three chains from $N_{\rm tot}
\geq 129$.} \label{fig:energy}
\end{figure}

\begin{figure}
\includegraphics[width = 8.5 cm]{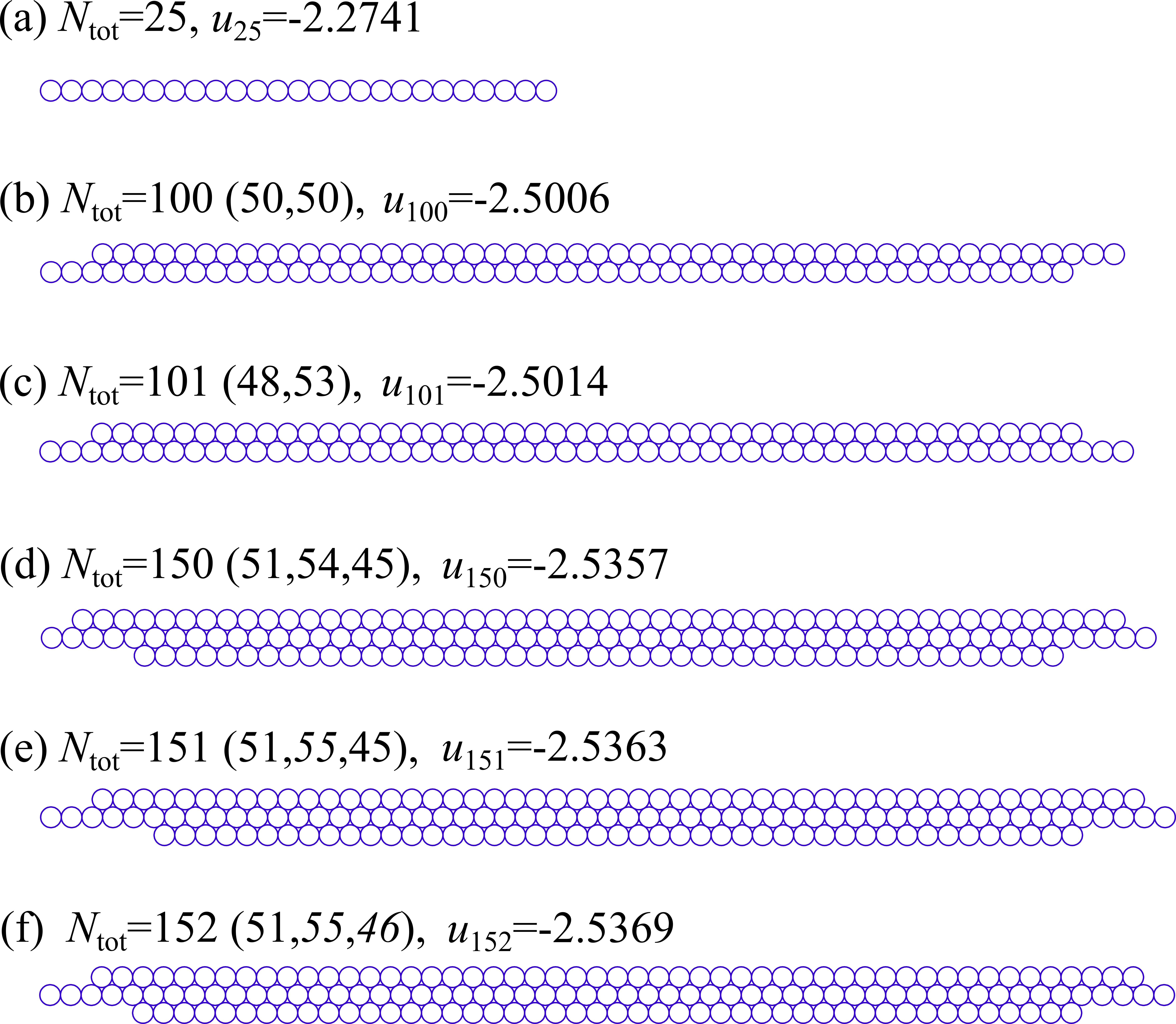}
\caption{ Typical ground state microstructures for different total
number of particles $N_{\rm tot}$. (a) One-chain configuration for
$N_{\rm tot}=25$. (b-f) Individual chain composition (from top to
bottom) are indicated as $(N_1,N_2)$, $(N_1,N_2,N_3)$ (from left
to right) for two- and -three-chain configurations, respectively.
(b,c) Two-chain configurations for $N_{\rm tot}=100,101$,
respectively. (d-f) Three-chain configurations for $N_{\rm
tot}=150,151,152$, respectively. } \label{fig:snaps}
\end{figure}

\section{Ground state structure of self-assembly}

\subsection{Energy minimization with genetic algorithm}

In that numerical part of the work, the reduced potential energy of
interaction $U^{tot}_{N_{\rm tot}}$ has been minimized 
by evolving transient configurations on a triangular lattice using genetic algorithm. 
In order to increase the chance of finding the ground state, 
we typically employ many independent populations of about $1000$ initial configurations consisting 
of individual particles positions. 
The particle occupation on the triangular lattice is mapped on an array of 0s and 1s,
meaning a particle is present at a certain position on lattice or not,
respectively. The evolution starts from a population of randomly
generated individuals (i.e., configurations), and consists of an iterative process 
of creation of the new generations. In each generation, the potential
energy of interaction $U^{tot}_{N}$ is evaluated for every individual in the population. 
The prime difference to usual deterministic minimization procedures is that multiple
individuals are evolved in each step in a stochastic manner: A number of
the best configurations are selected from the current population
and  modified (recombined and possibly randomly mutated) to form a new generation. 
The algorithm terminates when a prescribed maximum number of generations has been produced
and no more improvement of the best individual(s) is achieved.

\subsection{Numerical results}

The overall ground states at prescribed number of constitutive 
magnets $N_{tot}$ were also independently computed using a genetic algorithm.   
The resulting energy profiles for one chain, two and three touching chains  
are depicted in Fig.~\ref{fig:energy}. 
Corresponding illustrative microstructures are sketched in Fig. \ref{fig:snaps}.
The energy profile for the single chain structure stems from Eq.~\ref{eq.energy_intrachain}, 
whereas that for two touching chains was generated using
Eqs.~\ref{eq.energy_tot_2N} and~\ref{eq:U_2N+1^ass}, 
for even and uneven number of particles respectively. 
In concordance with the magnetic energy analysis of two-chain states, 
see Fig. \ref{fig:Utot_tails_equal_chain_size} and \ref{fig:energy_ass_2N+1}, 
minimization procedure based on genetic algorithm confirms, for $N_{tot}>26$
(see Fig. \ref{fig:energy}), 
that the ground states correspond indeed to either 
\begin{enumerate}[(i)]

\item
two touching equally long parallel chains shifted by two 
and a half beads for even $N_{tot}$, see Fig. \ref{fig:snaps}(b) for illustration,  

\item
or two symmetrically placed chains with $N+3$ and $N-2$ beads 
for uneven $N_{tot}=2N+1$, see Fig. \ref{fig:snaps}(c) for illustration.
\end{enumerate}

For $N_{tot}=26$, the energy per bead for the one-chain structure is 
$u_{26}^{(1-chain)}\simeq -2.2791$ whereas for the two-chain structure 
$u_{26}^{(2-chains)}\simeq -2.2781$ is found. 
For $N_{tot}=27$, we have $u_{27}^{(1-chain)}\simeq -2.2836$ against
$u_{27}^{(2-chains)}\simeq -2.2889$. 

Finally, we were able to locate the two- to three-chain transition at
$N_{tot}=129$ by means of genetic algorithm, cf. Fig. \ref{fig:energy}.
More specifically, at $N_{tot}=128$, the energy per bead for the two-chain structure is 
$u_{128}^{(2-chains)}\simeq -2.5182$ whereas for the three-chain structure 
$u_{128}^{(3-chains)}\simeq -2.5178$ is found.
At $N_{tot}=129$, we have 
$u_{129}^{(2-chains)}\simeq -2.5187$ against 
$u_{129}^{(3-chains)}\simeq -2.5189$.
Representative microstructures are depicted in Fig. \ref{fig:snaps}(d-f), 
where it can be clearly identified that the (long) mid-chain is always sandwiched by two shorter chains.

\section{Concluding remarks}

We have analyzed the self-assembly of magnetic spheres under strong magnetic field in two dimensions. 
Chains and assembly of chains constitute the typical microstructures of magnetic beads. 
A simple but highly instructive situation concerns a single magnet (i.e., a monomer) interacting 
with a magnetic chain. 
An effective attraction, even with a dimer, sets in at very short separation 
as a result of \textit{dipole-dipole correlations} and \textit{excluded volume} effects.
Nevertheless, a negative magnet-chain interaction energy is solely obtained for a long enough chain. 

It turns out that the case of two-chain ground state structures possesses strong symmetry.
This feature has allowed us to undertake exact analytical considerations about the 
behavior of the global potential of interaction.    
We have demonstrated that the attraction of parallel magnetic chains 
is essentially a near field effect (i.e., a range of the order of the magnetic monomer) 
and its strength strongly depends on chain length and relative position. 
In the far field limit, magnetic chains behave as super dipoles as expected. 
Besides, we have brought to light tail creation as a mechanism 
for reduction (more negative) of the cohesive energy. 
A general result, 
which holds independently of number of particles, 
is that minimal energy is achieved if the tails have
length of \textit{two and a half} beads.
In the case of even number of particles we
obtain a two-fold symmetric ribbon made up of two shifted like-sized chains. 
For uneven number of particles, we obtain hat-like
structures exhibiting a mirror symmetry.


Ground states of self-assembled magnets have been addressed by genetic algorithm.  
The resulting phase diagram (energy per bead vs the total number of beads $N_{tot}$) 
has the following characteristics:
\begin{itemize}

\item
At moderate number of beads $2 \leq N_{tot} \leq 26$, it is the \textit{one-chain} 
assembly that possesses the lowest energy.

\item
When $27 \leq N_{tot} \leq 128$, it is the \textit{two-chain} 
assembly that possesses the lowest energy.

\item
The \textit{two-} to \textit{three-chain} transition occurs at $N_{tot} = 129$.

\end{itemize}

Note that analog findings to the  three-chain structure were recently 
published ~\cite{Messina_PREReply_2015} in three dimensions, 
where straight long magnetic chains  develop a rodlike structure with a narrow rectangular section.

\begin{acknowledgments}

The  authors  acknowledge  financial  support  from  the bilateral
Franco-Serbian  PHC  Pavle  Savic  2014/15  program (No. 32135NJ).
I.S.  acknowledges  support  received  from the Serbian Ministry
of Education and Science (Grants No. ON171017 and No. III45018).
Numerical simulations were run on the PARADOX supercomputing
facility at the Scientific Computing Laboratory of the Institute
of Physics Belgrade.

\end{acknowledgments}

\end{document}